\documentclass[fleqn,12pt]{article}
\usepackage{cite,amssymb}

\def\noi{\noindent}

\makeatletter
\renewcommand{\section}{\@startsection{section}{1}{0pt}%
        {-3.5ex plus -1ex minus -.2ex}{2.3ex plus .2ex}%
        {\large\bf\protect\raggedright}}

\renewcommand{\subsection}{\@startsection{subsection}{2}{0pt}%
        {-3ex plus -1ex minus -.2ex}{1.4ex plus .2ex}%
        {\normalsize\bf\protect\raggedright}}

\renewcommand{\thesubsubsection}%
        {\arabic{section}.\arabic{subsection}.\arabic{subsubsection}.}

\renewcommand{\@oddhead}{\raisebox{0pt}[\headheight][0pt]{%
   \vbox{\hbox to\textwidth{\rightmark \hfil \rm \thepage \strut}\hrule}}}
\renewcommand{\@evenhead}{\raisebox{0pt}[\headheight][0pt]{%
   \vbox{\hbox to\textwidth{\thepage \hfil \leftmark \strut}\hrule}}}

\makeatother

\newcommand{\Author}[2]{\noi{\large\bf #1}\\[2ex]\noindent{\it #2}\\}

\newcommand{\Abstract}[1]{\vskip 2mm \begin{center}
        \parbox{16.4cm}{\small\noi #1} \end{center}\medskip}

\newcommand{\sect}[1]{Sec.\,#1}

\def\nqq{\hspace*{-2em}}
\def\nhq{\hspace*{-0.5em}}

\def\cm{\hspace*{1cm}}

\def\ten#1{\mbox{$\,\times\, 10^{#1}$}}

\def\Jl#1#2{{\it #1\/} {\bf #2},\ }

\def\ApJ#1 {\Jl{Astroph. J.}{#1}}
\def\CQG#1 {\Jl{Class. Quantum Grav.}{#1}}
\def\DAN#1 {\Jl{Dokl. AN SSSR}{#1}}
\def\GC#1 {\Jl{Grav. \& Cosmol.}{#1}}
\def\GRG#1 {\Jl{Gen. Rel. Grav.}{#1}}
\def\JETF#1 {\Jl{Zh. Eksp. Teor. Fiz.}{#1}}
\def\JETP#1 {\Jl{Sov. Phys. JETP}{#1}}
\def\JHEP#1 {\Jl{JHEP}{#1}}
\def\JMP#1 {\Jl{J. Math. Phys.}{#1}}
\def\NPB#1 {\Jl{Nucl. Phys.}{B\ #1}}
\def\NP#1 {\Jl{Nucl. Phys.}{#1}}
\def\PLA#1 {\Jl{Phys. Lett.}{#1A}}
\def\PLB#1 {\Jl{Phys. Lett.}{#1B}}
\def\PRD#1 {\Jl{Phys. Rev.}{D\ #1}}
\def\PRL#1 {\Jl{Phys. Rev. Lett.}{#1}}

\newcommand{\eqsection}{\makeatletter
    \@addtoreset{equation}{section}
    \renewcommand{\theequation}{\arabic{section}.\arabic{equation}}
    \makeatother}
\def\al{&\nhq}
\def\lal{&&\nqq {}}
\def\eq{Eq.\,}

\def\beq{\begin{equation}}
\def\eeq{\end{equation}}
\def\bear{\begin{eqnarray}}
\def\bearr{\begin{eqnarray} \lal}
\def\ear{\end{eqnarray}}
\def\earn{\nonumber \end{eqnarray}}
\def\nn{\nonumber\\ {}}

\def\yy{\\[5pt] {}}

\def\eql{\al =\al}

\def\tst{\textstyle}

\def\fract#1#2{{\tst\frac{#1}{#2}}}

\def\half{{\fract{1}{2}}}

\def\e{{\,\rm e}}

\def\const{{\rm const}}
\def\eps{\varepsilon}
\newcommand{\vars}[1]{\left\{\begin{array}{ll}#1\end{array}\right.}

\def\mn{_{\mu\nu}}
\def\MN{^{\mu\nu}}

\begin{document}
\thispagestyle{empty}

\noi
{\Large\bf Possible variations of the fine structure constant $\alpha$\yy
    and their metrological significance}

\bigskip

\Author{\normalsize K.A. Bronnikov and S.A. Kononogov}
    {VNIIMS, 46 Ozyornaya St., Moscow 119361, Russia}

\Abstract{We briefly review the recent experimental results on possible
      variations of the fine structure constant $\alpha$ on the
      cosmological time scale and its position dependence. We outline
      the theoretical grounds for the assumption that $\alpha$ might be
      variable, mention some phenomenological models incorporating a
      variable $\alpha$ into the context of modern cosmology and discuss
      the significance of possible $\alpha$ variations for theoretical
      and practical metrology.}

\section{Introduction}

  The idea that fundamental physical constants (FPC), including the Planck
  constant, the speed of light, stable particle masses and the coupling
  constants of the basic interactions (above all, Newton's gravitational
  constant $G$ and the fine structure constant $\alpha$) may be variable,
  was put forward for the first time by Dirac in 1937, soon after the advent
  of Friedmann's expanding Universe models (1922) and their confirmation by
  Hubble (1929). In the subsequent decades, the possible FPC variability in
  the course of cosmological evolution as well as their dependence on
  position in space, on the magnitude of physical fields or on spatial and
  energy scales occupied quite a prominent place in theoretical physics and
  cosmology, including many works of the VNIIMS theoretical group,
  Refs.\,\cite{v1,v2,v3,v4,v5,v6,v7,v8,v9,1,3} and others. In addition to the
  above-mentioned FPC, there appeared other ``candidate variables'' such as,
  for instance, the cosmological constant and the coupling constants of the
  weak and strong interactions. The basic motivation for such studies has
  been, and remains to be, the idea of unity of physical interactions. If
  the conjectured unified interaction splits in different physical
  conditions into specific interactions known from the experiment, the FPCs
  that characterize these interactions should depend on the circumstances of
  their manifestation, above all, on the spatial and energy scales and on
  the fundamental field intensities (a recent detailed discussion may be
  found in Ref.\,\cite{1}).

  Thus, it is quite common to believe that unification of all physical
  interactions requires a number of additional (so-called internal)
  space-time dimensions. At present, there is no well established way of
  transition from multidimensional physics to the conventional
  four-dimensional picture. It is, however, clear that such a transition
  requires some procedures (such as dimensional reduction and
  compactification) which connect the four-dimensional effective constants
  with true multidimensional constants and the characteristics of extra
  dimensions. Thus, their size may suffer a non-trivial evolution in time
  and may depend on the position in external space.

  As to observational data on FPC variation, there have been very few positive
  results till the recent years. Thus, some researchers pointed toward a
  possible $G$ variability according to astronomical observations, but, to
  our knowledge, neither of these results were confirmed afterwards (see
  Ref.\,\cite{1} for detail). However, the very end of the 20th century was
  marked with the advent of observational evidence \cite{2} on the
  variability of the electromagnetic coupling constant, the fine structure
  constant
\bear                                                      \label{1}
    \alpha \eql e^2/(\hbar c) \quad \mbox{(Gaussian units), or}
\nn
    \alpha \eql e^2/(4\pi \eps_0 \hbar c)\quad \mbox{(SI units)},
\ear
  where $e$ is the electron charge, $c$ is the speed of light in vacuum and
  $\eps_0$ is the permittivity of free space. This and subsequent works have
  made the subject of FPC variability, and above all the variability of
  $\alpha$, highly topical.  This subject is no longer regarded purely
  theoretical, and nowadays it rapidly increases its experimental basis.

  The electromagnetic interaction, whose intensity is characterized by
  the constant $\alpha$, is of primary importance for the macroscopic
  structure of matter and in an overwhelming majority of the observed
  phenomena. It is for this reason that a possible variability of $\alpha$
  would lead to variability of the majority of existing references of
  physical quantities, including such basic references as those of length,
  mass, time/frequency.

  The constant $\alpha$, whose CODATA-recommended value is
\[
      \alpha = 7.297362533(27)\ten{-3} \approx 1/137.03599
\]
  (the figures in parentheses mean the uncertainty in the last two meaningful
  digits), is one of the most precisely measured FPC: the relative
  uncertainty is about $3.7 \ten{-9}$. Moreover, its dimensionless nature
  makes it independent of any conventions or systems of physical units.

  In what follows, we will briefly describe the basic experimental and
  observational data on the possible $\alpha$ variability and outline the
  main characteristics of theoretical models which, to one or other extent,
  conform to the observations. In the last section, we formulate some
  conclusions and discuss the significance of this research area for
  theoretical and practical metrology.

\section {Empirical data on $\alpha$ variability}

   A great number of relevant observations and experiments are described and
   discussed in the comprehensive review \cite{Uzan03} and in the shorter
   articles \cite{Ksh03,37}. In this section, we mention only some of the
   data presented there, seeming to be the most convincing, and add some
   estimates and inferences of the last two years.

\subsection {Laboratory data}

   The laboratory bounds on $\alpha$ variability are obtained by comparing
   the long-term behaviour of oscillators with frequencies depending
   on $\alpha$ in different ways. In practice, such a
   comparison leads to experimental bounds on certain combinations of FPCs.

   To our knowledge, the first comparisons of this kind were performed
   in 1974-76 by two groups independently. Turneaure and Stein \cite{TS76},
   at Stanford University, used superconducting microwave cavities with a
   quality factor of about $4\ten{10}$ and a resonant frequency $\sim 8.6$
   GHz. The frequencies of cavity-controlled oscillators were compared over
   a period of 12 days with a caesium-beam atomic clock, and the relative
   drift was $(-0.4 \pm 3.4)\ten{-14}/{\rm day}$. This led to the conclusion
   \cite{TS76} that a possible annual variation of the product
\[
      g_p \cdot(m_e/M)\cdot \alpha
\]
   was less than $4.1\ten{-12}$, where $g_p$ is the proton gyromagnetic
   ratio, $m_e$ is the electron mass and $M$ is the Cs nuclear mass.

   Kolosnitsyn et al. \cite{3}, at VNIIFTRI, the Russian
   time-frequency standard keeper, compared the behaviour of molecular
   clocks operating with ammonia molecular beams and a group of caesium
   frequency standards. The caesium standard formed a scale for measuring
   the frequency of the ammonia generators. The results of measurements
   for six years (1964-1969) led to a number of estimates for
   $\dot\alpha/\alpha$, all of them constraining this variation by a few
   units times $10^{-11}$; the best estimate was
\[
    |\dot{\alpha}/\alpha| \leq 1.1\ten {-11} \,{\rm yr}^{-1}.
\]

   Other early results on the stability of $\alpha$ and $e$ are described
   in Petley's book \cite{Petley}.

   The recent estimates of possible $\alpha$ variations have diminished the
   uncertainties by 4 to 5 orders of magnitude as compared with \cite{3} due
   to a considerably increased stability of frequency standards. Thus,
   according to \cite{4}, a comparison of hyperfine transition frequencies
   in $^{87}$Rb and $^{133}$Cs over a period of about 4 years has shown that
\[
    d\ln(\nu_{\rm Rb}/\nu_{\rm Cs})/
    d t = (0.2 \pm 7.0 )\times 10^{-16}\,{\rm yr}^{-1}
\]
   at 1$\sigma$. Neglecting possible changes in the the weak and strong
   interaction constants and thus in the nuclear magnetic moments, it leads
   to
\beq
    \dot\alpha/\alpha =
            (-0.4\pm 16)\times 10^{-16}\,{\rm yr}^{-1}.
\eeq
   where $\dot{\alpha} = d\alpha/dt$. Another experiment \cite{5} compared
   an electric quadrupole transition in $^{199}$Hg$^+$ to the ground-state
   hyperfine splitting of $^{133}$Cs over 3 years, showing that
\[
    |d\ln(\nu_{\rm Hg}/\nu_{\rm Cs})/d t| <7.0\times
                    10^{-15}\,{\rm yr}^{-1}
\]
   and leading to
\beq
    \left|\dot\alpha/\alpha\right| < 1.2\times 10^{-15}\,{\rm yr}^{-1}
\eeq
   under the assumption that the gyromagnetic factor $g_{\rm Cs}$ and
   $m_{\rm e}/m_{\rm p}$ are invariable.

   Ref.\,\cite{6} reported a comparison of the absolute $1S-2S$ transition
   in atomic hydrogen to the ground state of caesium, which, combined with
   the results of Refs. \cite{3,5}, yielded the constraint
\beq
    \dot\alpha/\alpha = (-0.9\pm2.9)\times10^{-16}\,{\rm  yr}^{-1}.
\eeq

   A comparison \cite{7} of optical transitions in $^{171}$Yb$^+$ with a
   caesium standard for 2.8 years has shown that
   $d\ln(\nu_{\rm Yb}/\nu_{\rm Cs})/d t = (-1.2 \pm 4.4) \times
   10^{-15}\,{\rm yr}^{-1}$ which leads to
\beq
    \dot\alpha/\alpha = (-0.3\pm2.0)\times10^{-15}\,{\rm yr}^{-1}.
\eeq

   The laboratory measurements thus lead to results which, as follows from
   Ref.\,\cite{6}, may be combined to yield very hard constraints for
   separate basic constants including $\alpha$. The above estimates show
   that the allowed variations of $\alpha$ in the modern epoch are bounded
   to a few units of $10^{-16}$ per year.

   Further progress in the near future is expected on the basis of improved
   frequency standards as well as development of new methods of atomic
   experiments. One such method \cite{7a} is based on crossing of atomic
   levels in two-electron highly charged ions of different atomic numbers
   $Z$. It is claimed that the effect of possible variation of $\alpha$
   becomes strongly (by a factor of about 1000) enhanced when studied near
   such crossing points. Another method \cite{7b} is to use atomic
   transitions between narrow lines that have an enhanced sensitivity to a
   possible variation of $\alpha$. The authors present a number of such
   transitions and claim that their method effectively suppresses the
   systematic sources of uncertainty that are unavoidable in conventional
   high-resolution spectroscopic measurements.

   An improvement of at least an order of magnitude in current $\dot\alpha$
   estimates can be achieved in space experiments planned for 2008-2009 within
   the PHARAO/ACES Project of the European Space Agency \cite{ACES-05,ACES-01}.
   PHARAO, the ``Projet d`Horloge Atomique par Refroidissement d`Atomes
   en Orbite'', which has started in 1993, combines laser cooling
   techniques and a microgravity environment in a satellite orbit, allowing the
   development of space clocks with unprecedented performance. Its
   declared objectives are, among others \cite{ACES-05}, (i) to achieve time
   and frequency transfer with stability better than $10^{-16}$ and
   (ii) to perform fundamental physics tests. Its detailed description
   can be found in \cite{ACES-01}.

   One can also mention quite a different type of $\alpha$ variability
   predicted by the electroweak theory which concerns strengthening of
   the electromagnetic coupling with increasing momentum transfer $Q^2$ in
   particle interactions at energies approaching the electroweak unification
   energy \cite{Levine, Tobar}. This effect has been confirmed experimentally
   \cite{Levine} by analyzing the processes $e^+e^- \to \mu^+\mu^-$ and
   $e^+e^- \to \e^+\e^-\mu^+\mu^-$ at the TRISTAN $e^+e^-$ collider at KEK
   (Japan). The data were accumulated at an average center-of-mass energy
   of 57.77 GeV, and $\alpha$ was measured to change from its known
   value of $\alpha^{-1} \approx 137.0$ at $Q^2 = 0$ to
\[
   \alpha^{-1} = 128.5 \pm 1.8\ {\rm (stat)} \pm 0.7\ {\rm (syst)}
   \quad\ {\rm at} \quad\ Q^2 = (57.77\ {\rm GeV}/c)^2.
\]
   The latter variation is physically interpreted as a result of quantum
   vacuum properties (see, e.g., \cite{Tobar} and references therein) and
   manifests itself at energies much higher than those relevant to atomic
   spectra. It has therefore nothing to do with temporal and spatial
   variations of $\alpha$ at astrophysical and cosmological scales.

\subsection{Geochemical data}

   Very hard constraints on variations of $\alpha$ have been obtained from
   studies of the so-called Oklo phenomenon combined with data on the
   lifetimes of long-lived radioactive isotopes.

   The Oklo phenomenon is a natural nuclear reactor that operated for about
   200,000 years approximately two billion years ago, which, regarding
   cosmological observations, corresponds to redshifts around $z \sim 0.14$.
   The products of its operation were discovered in 1972 at Oklo uranium
   mine in Gabon, West Africa. The isotopic abundances of the yields bears
   information on the nuclear reaction rates at that time, which in turn
   depended on the current values of $\alpha$. One of the key quantities
   measured is the ratio ${}^{149}_{62}{\rm Sm}/{}^{147}_{62}{\rm Sm}$ of
   two light isotopes of samarium which are not fission products.  This
   ratio is of the order of 0.9 in normal samarium, but is about 0.02 in
   Oklo ores. Such a low value is explained by depletion of
   ${}^{149}_{62}{\rm Sm}$ by thermal neutrons to which it was exposed while
   the reactor was active.  The capture cross section of a thermal neutron
   by $^{149}{\rm Sm}$, i.e., the reaction
\[
    ^{149}{\rm Sm} + n \to  ^{150}{\rm Sm} + \gamma
\]
   has a resonant energy $E_r \simeq 0.0973$~eV, which is a consequence
   of a nearly cancellation between electromagnetic and strong
   interaction forces \cite{8}. A detailed analysis has shown \cite{9} that,
   assuming that $E_r$ varied only due to an $\alpha$ dependence of the
   electromagnetic energy, the following constraint is valid:
\beq
         \Delta\alpha/\alpha = (0.15\pm1.05)\times 10^{-7}
\eeq
   at $2\sigma$ level. The accuracy of the method is explained by the tiny
   value of the resonant energy $E_r\sim 0.1$ eV compared to its
   sensitivity to a variation of $\alpha$, $d E_r/d\ln\alpha \sim - 1$ MeV,
   so that the expected variations are smaller than $0.1\,{\rm eV}/1\,{\rm
   MeV}\sim 10^{-7}$.

   It was later noticed \cite{10} that two ranges of variations could be
   compatible with the Oklo data:
\beq
    \Delta\alpha/\alpha = (-0.8\pm 1.0)\times 10^{-8},\qquad
    \Delta\alpha/\alpha = (8.8\pm 0.7)\times  10^{-8},
\eeq
   but the second range is hardly compatible with the isotopic ratio of
   gadolinium. Note that the first range, assuming that the time variations
   of $\alpha$ (if any) occur uniformly, is translated to the variation rate
\beq
    \dot{\alpha}/\alpha                                 \label{e7}
            = (-0.4 + 0.5) \ten{-17}\ {\rm yr}^{-1},
\eeq
   which is apparently the tightest of the existing constraints.

   Some estimates have been recently obtained under the additional
   assumption that the low energy neutron spectrum is well described by a
   Maxwell-Boltzmann distribution \cite{11,12,13}. Taking into account the
   possible variation of the strange quark mass $m_{\rm s}$ along with the
   assumption that all fundamental couplings vary independently led \cite{13}
   to the allowed variation $\Delta\alpha/\alpha<(1-5)\times10^{-10}$, even
   more stringent than above.

   Another result \cite{14}, free of poorly grounded assumptions,
   seems more realistic:
\beq                                                            \label{e8}
   \Delta \alpha/\alpha = (4.5 + 1) \times 10^{-8}, \qquad
   \dot\alpha/\alpha = (2.25 + 0.5) \times 10^{-17}\ {\rm yr}^{-1} .
\eeq

   The most recent $\dot{\alpha}$ estimates from the Oklo phenomenon
   \cite{Petrov05} have been made on the basis of modern methods of reactor
   physics, a detailed computer model of zone RZ2 of the Oklo reactor and
   full-scale calculations using two different Monte Carlo codes: the
   Russian code MCU REA and the worldwide known code MCNP (USA).
   Both codes have produced close results. It was claimed that many details
   of these calculations differed from the previous work (e.g., the averaged
   cross-section of Sm and its dependence on the shift of resonance
   position due to variation of fundamental constants). Still no evidence on
   $\alpha$ variations was found, with the resulting constraint
\[
   -4 \ten{-17} {\rm yr}^{-1} < (d \alpha/dt)/\alpha
                        < 3\ten{-17} {\rm yr}^{-1}.
\]
   A further improvement in the accuracy of these limits is promised.

   Some conclusions have also been made by analyzing the readioactive decay
   of nuclides with known $\alpha$ dependence of the decay rate. The best
   constraint was obtained from beta decay of rhenium into osmium with
   electron emission: as noted by Peebles and Dicke \cite{15}, an extremely
   low value of the decay energy, about 2.5 keV, makes it very sensitive to
   $\alpha$ variations. Still the constraints obtained are much weaker than
   (\ref{e7}) or (\ref{e8}).

   Laboratory studies of meteorites have led to the result \cite{16}
\beq
    \Delta\alpha/\alpha = (8 \pm 16)\ten{-7}                 \label{e9}
\eeq
   for the recent 4.5 billion years, which corresponds to $z\simeq 0.45$.
   Translated to the changing rate (again assuming its uniformity) gives
   approximately
\beq
    \dot\alpha/\alpha = (2 \pm 4)\ten{-16}/{\rm yr}.         \label{e10}
\eeq

   A shortcoming of this method is its indirect nature: the Re/Os ratio
   is measured in iron meteorites whose age is not determined directly.
   Besides, the constraint (\ref{e9}) only concerns the mean value of
   $\alpha$ for 4.5 billion years rather than its instant variation rate.

\subsection{Astrophysical data}

   The first and not very confident observational indication on a possible
   variability of $\alpha$ has appeared from studies of remote quasars.
   The observed quasar absorption spectra were compared with the
   corresponding laboratory spectra. The details of the method are described
   in Refs.\,\cite{2,17,18,19,20} and others.

   One of the recent results \cite{18} has been obtained by analyzing the
   lines of SiIV in 15 systems and improves the previous estimates by a
   factor of three:
\beq
   \Delta\alpha/\alpha = (0.15\pm 0.43)\ten{-5},\qquad 1.59\leq z\leq 2.92.
\eeq
   This result does not confirm $\alpha$ variations.

   The many-multiplet method of Webb et al. \cite{2, 19} rests on a
   comparison of a few lines in different samples of sources: one compares
   the shifts of lines which are sensitive and insensitive to $\alpha$
   variations. A recent result [20] obtained from a study of 128 systems
   in the range $0.5 < z < 3$ points at lower values of $\alpha$ in the past:
\beq
     \Delta\alpha/\alpha=(-0.54 \pm 0.12)\ten{-5},\qquad 0.5\leq z\leq 3,
\eeq
   thus confirming the previous conclusions \cite{2, 19}. The authors did
   not find systematic errors.

   There are, however, results directly opposite to these. Thus, according
   to \cite{21},
\beq
        \Delta\alpha/\alpha=(-0.1 \pm 1.7)\ten{-6},\qquad z=1.15
\eeq
   (from an analysis of absorption lines of iron of a single quasar).
   Similar results have been found for a system of absorption lines of the
   quasar Q 1101-264 \cite{22} with the redshift $z=1.839$:
\beq
       \Delta\alpha/\alpha = (2.4 \pm 3.8)\ten{-6},\qquad z=1.839.
\eeq
   A unified sample of Fe II lines gave $\Delta\alpha/\alpha =
   (- 0.4\pm 1.5)\ten {-6}$ for two systems with $z=1.15$ and $1.839$.

   The use of radio and millimeter quasar absorption spectra gave
   an independent estimate (unfortunately, for comparatively small redshifts
   only)
\beq
       |\Delta\alpha/\alpha| < 8.5\ten{-6},\qquad  z = 0.25 - 0.68.
\eeq

   As noted by Barrow \cite{barr05}, the same quasar absorption spectra
   analysis leads to an upper bound on spatial variations of $\alpha$ like
   $|\Delta \alpha|/\alpha < 3 \ten{-6}$ at 3 Gpc distance scale because of
   a wide distribution of the target absorption systems over the sky.

   An alternative method is to study emission rather than absorption
   spectra. There are very few such estimates since this method, being
   sufficiently simple and direct, is less sensitive and is harder to apply
   to sources with large redshifts. Recent measurements of strong emission
   lines of O III in a sample of 165 quasars have led to the result \cite{24}
\beq                                                          \label{e16}
     \Delta\alpha/\alpha = (1.2 \pm 0.7)\ten{-4},\qquad z = 0.16 - 0.8.
\eeq
   Essential improvements in emission line measurements and analysis is
   anticipated in the near future \cite{Grupe05}, partly owing to the use of
   many-multiplet methods already employed in the absorption spectra
   analysis.

   The estimate (\ref{e16}) as well as tentative estimates of the Cambridge
   group (England) \cite{25} point at higher values of $\alpha$ in the past
   whereas the results related to absorption lines give preference to
   smaller $\alpha$ in the past. It appears tempting to conclude that this
   apparent discrepancy may be explained by some unidentified systematic
   errors. Arguments in favour of this viewpoint have been put forward by
   Bandeira and Corbelli \cite{BaCo05}, who found that at least the
   absorption line analysis could be subject to systematic effects, related
   to estimation of different sets of atomic transitions at different
   redshifts, and these hidden correlations thus could mimic a variable
   $\alpha$.

   Another, ``optimistic'' viewpoint is, however, possible \cite{25}, that
   this is simply an indication of spatial variations of $\alpha$ since the
   emission and absorption methods are sensitive to the values of $\alpha$
   in drastically different environments.

   One can conclude that the present astrophysical data on $\alpha$
   variation are rather contradictory and need further verification and
   improvement.

\subsection {Cosmological data}

   The observed anisotropy of the cosmic microwave background (CMB) and the
   abundance of light elements formed in primordial nucleosynthesis (PNS)
   constrain the variations of $\alpha$ on the cosmological scale on the
   level $10^{-2}$. These constraints concern much greater redshifts $z$
   than any others: $z \sim 1000$ for CMB and $z \sim 10^{10}$ for PNS.

   Variations of $\alpha$ change the Thomson scattering cross-section and
   hence the transparency of the medium and finally the recombination time.

   A recent analysis of the WMAP (Wilkinson Microwave Anisotropy Probe)
   data on CMB anisotropy has led to the result \cite{26}
\beq
    \Delta\alpha/\alpha = (-1.5 \pm 3.5) \ten{-2},         \label{CMB}
\eeq
   for $z\sim 10^3$. It is, however, well known that a similar effect on CMB
   anisotropy could be due to variations of the gravitational constant
   \cite{27}. Besides, consistent constraints on $\alpha$ variations of the
   order of $1\%$ could only be obtained from CMB analysis combined with
   independent data on the cosmological parameters characterizing the
   expansion of the Universe at times close to the recombination epoch.

   Barrow \cite{barr05} used the observed CMB isotropy at large angular
   scales to derive strong observational limits on any possible large-scale
   spatial variation in the values of $\alpha$ and other FPC (the electron
   to proton mass ratio and the Newtonian gravitational constant) assuming
   that their space-time evolution is driven by a scalar field. The
   constraints are strongly model-dependent. Thus, large-scale spatial
   fluctuations of $\alpha$ are bounded by 2\ten{-9} in the BSBM theory
   (see below) and by 1.2\ten{-8} in the varying speed of light theories.
   These derived bounds are significantly stronger than any others,
   obtainable by direct observations of astrophysical objects. It should be
   noted, however, that these bounds concern variations between mutually
   remote regions of space with approximately equal physical conditions and
   do not apply to the possible discrepancy between the results of emission
   and absorption spectra of quasars (see the end of the previous section):
   in the latter case, the emitters and absorbers are apparently
   characterized by drastically different gravitational fields, matter
   densities and temperatures.

   The PNS theory predicts light element formation in the early Universe,
   and their resulting abundances depend on a delicate balance between the
   Universe expansion and the weak interaction strength which in turn
   determines the proton to neutron number ratio at the beginning of PNS.
   Ultimately, the predicted light element abundances depend on a number of
   fundamental constants, $\alpha$ being only one of them.

   A recent study incorporating seven parameters \cite{28} has yielded
\beq                                                            \label{PNS}
    \Delta\alpha/\alpha = (6\pm 4)\ten{-4}, \qquad z \sim 10^{10}.
\eeq

   A considerable improvement of the above estimates is expected with the
   appearance and analysis of new observational data from the WMAP and
   Planck Surveyor satellites.

   According to Ref.\,\cite{25}, the CMB and PNS analysis on the whole
   favours slightly smaller values of $\alpha$ in the past compared to its
   modern value. The absence of variations also remains admissible on the
   level of $2\sigma$.

   The bounds (\ref{CMB}) and (\ref{PNS}) cannot be directly translated into
   estimates of the admissible current values of $\dot\alpha$ since, to do
   so, one has to assume a certain dependence $\alpha(t)$ which is not only
   different in different theories, but even in specific models of a given
   theory; see examples in \sect 3.

\subsection{Comparing data of different origin}

   The above discussion of possible FPC variability combined observational
   evidence from quasar spectra with Solar system data and laboratory
   constraints. Most of the studies implicitly assume that local and
   cosmological observations are directly comparable. This is, however, a
   strong assumption which may prove to be wrong. If a given FPC, say,
   $\alpha$, depends on a scalar field $\phi$, a slowly varying cosmological
   background of $\phi$ may substantially differ from its local values in
   a specific galaxy, stellar cluster or another gravitationally bound
   object. One can freely imagine, for instance, that the Oklo data only
   reflect the value of a stationary galactic scalar field whereas the
   quasar data feel cosmological effects.

   Shaw and Barrow \cite{shaw05} have found that it is probably not the
   case, and local variations of the ``constants'' are able to track their
   global variations. A construction involving matched asymptotic expansions
   within a sufficiently wide scalar field model is analyzed assuming that
   the scalar (dilaton) field is only weakly coupled to gravity and has a
   negligible effect on the background space-time geometry; a local
   inhomogeneity was characterized by the Tolman-Bondi class of spherically
   symmetric metrics describing the evolution of a dust cloud in the
   presence of a cosmological constant. It was concluded that local
   virialization does not stabilize the value of the dilaton and protect it
   from any global cosmological variation. This indicates that local
   terrestrial and Solar-system experiments really measure the effects of
   varying ``constants'' of Nature occurring on cosmological scales to
   computable precision \cite{shaw05}.

   Somewhat different results have been obtained by Mota and Barrow
   \cite{mota04a,mota04b}.
   They studied the space and time evolution of $\alpha$ using the BSBM
   varying-$\alpha$ theory (see \sect 3.1) and a spherical collapse model
   for cosmological structure formation, considered as a perturbation to a
   homogeneous and isotropic cosmological background. Different assumptions
   were used on the dark energy equation of state and on the coupling of
   $\alpha$ to the matter fields. It was found that, independently of the
   model of structure formation one considers, there is always a difference
   between the values of alpha in a virialized overdensity and in the
   background.  In some models, especially at low redshifts, the difference
   depends on the time when virialization has occurred and the equation of
   state of the dark energy. At low redshifts, when the dark energy starts
   to dominate the cosmological expansion, the difference between alpha in a
   cluster and in the background grows.

   According to these studies, it may happen that $\alpha$
   and $\dot\alpha$ vary from galaxy to galaxy even at the same redshift,
   but the evolution of $\alpha$ is the same in different sub-galactic
   stellar systems belonging to a single galaxy, or even in galaxies
   belonging to the same galaxy cluster \cite{mota-let}. It is, however,
   clear that the problem should be studied more thoroughly in a wider
   range of models.

   More generally, one can conclude that the whole set of modern
   experimental and observational data leads to very stringent limits of
   putative $\alpha$ variations but leaves open the question of their real
   existence. Even more uncertainty follows from the possible difference
   between local and global $\alpha$ variations and its model dependence.

\section {Theoretical models that predict varying $\alpha$}

   Despite the weak experimental status of varying $\alpha$, there have been
   a great number of theoretical studies in the recent years, treating
   $\alpha$ as a function of certain physical fields which, in general,
   change from point to point in four-dimensional space-time.

   It should be noted that the researchers themselves stress a tentative
   nature of such constructions. Thus, according to \cite{25}, ``any
   model compatible with all empirical data at a given time, is certainly
   wrong since at any time some of the current data are erroneous''. It is
   therefore useless to build artificial models with a large number of free
   parameters: it is highly probable that such a model will be incompatible
   with the newest data as soon as its description is published.

   In studies dealing with multidimensional theories, such as Kaluza-Klein
   theories and different versions of string theory (regarded at present
   as the most promising unification theories), the dimensionless constants
   of four-dimensional physics are shown to be dynamical quantities
   \cite{30,31}. However, the specific form of such dynamics not only
   depends on the version of the Kaluza-Klein or string theory, but also on
   the compactification scheme and the form of dilatonic coupling adopted
   within the chosen version of the theory.

   Accordingly, at the present stage, comparatively simple and natural
   phenomenological models are preferable, which, on the one hand, should
   conform to the idea of unity of all physical interactions and, on the
   other, should not only be able to predict the variability of $\alpha$ but
   also suggest solutions to other problems of modern cosmology, such as the
   cosmological constant problem and the dark matter and dark energy
   problems. We will briefly outline two such models and mention some other
   approaches.

\subsection{Bekenstein-Sandvik-Barrow-Magueijo (BSBM) theories }

   This class of theories \cite{32,33} assumes constant values of the speed
   of light and Planck's constant; accordingly, a variability of $\alpha$ is
   achieved due to variability of the electron charge $e$ or the
   permittivity of vacuum. Thus instead of $e=\const$ ane assumes
   $e = e_0 \eps(x)$, where $e_0 = \const$ while $\eps$ is a dimensionless
   scalar field depending on the space-time coordinates. Such theories
   preserve the usual properties of local gauge and Lorentz invariance and
   causality. In these theories, the conventional expression for the
   electromagnetic field tensor in terms of the 4-potential $A_\mu$ is
   replaced by
\beq
       F_{\mu\nu} = (1/\eps)[(\eps A_\nu)_{,\mu} - (\eps A_\mu)_{,\nu}],
\eeq
   which takes its usual form if $\eps = \const$. The electromagnetic field
   action has its usual form while the scalar field dynamics is described by
   the usual kinetic term in the Lagrangian (written in the form
   $\eps_{,\mu} \eps^{,\mu}/\eps^2$). The theories of this class differ from
   one another in assumptions on how the $\eps$ field is coupled to the rest
   of matter.

   The cosmological models obtained in these theories predict time
   variations of $\alpha$ (e.g., $\sim \ln t$ in the matter-dominated phase
   in a spatially flat cosmology) and spatial variations of $\alpha$
   whose nature and evolution depends on the dark energy equation of state
   and the coupling of $\alpha$ to the matter fields \cite{mota04a,mota04b}.
   Inclusion of the effects of inhomogeneity, so that the dependence
   $\alpha(t)$ differs in the cosmological background and in virialized
   overdensities and may naturally lead to no observable local time
   variations of $\alpha$ on Earth and in our Galaxy even though such
   variations can be significant on quasar scales.

\subsection {Fujii's scalar-tensor theory}

   The theory suggested by Fujii \cite{34} contains, in addition to the
   metric tensor, two scalar fields. One of them  ($\sigma$) is a dilatonic
   Brans-Dicke type field with the potential
    $V(\sigma) = \Lambda \e^{-4\zeta \sigma}$
   (in the Einstein conformal frame, or picture), where $\Lambda$
   is a constant playing the role of a cosmological constant in the Jordan
   picture; $\zeta$ is the so-called non-minimal coupling constant
   characterizing a coupling between the $\sigma$ field and the metric in
   the Jordan picture. Another scalar field $\phi$ has a conventional
   kinetic term and interacts with $\sigma$ in such a way that their common
   potential in Einstein's picture has the form
\beq
     V(\phi,\sigma) = \e^{-4\zeta\sigma}
        [\Lambda + \half m^2\phi^2(1+\gamma \sin(k\sigma))],
\eeq
   where $m,\ \gamma,\ k$ are constants (approximately of order unity in the
   Planck system of units).

   Considering Jordan's picture as a physical one, the author studies the
   modification of standard quantum electrodynamics in this theory caused by
   the scalar fields and concludes that the effective electric charges are
   $\sigma$-dependent. The time dependence of $\sigma$ itself is affected by
   the $\phi$ field which leads to decaying oscillations of $\sigma$ in the
   modern epoch.

   A cosmological model built in the framework of this theory predicts
   comparatively rapid changes of $\alpha$ in the epoch corresponding to
   $z >1$ (but close enough to 1) and much slower changes in the modern
   epoch. This allows one to reconcile the strong bounds related to the Oklo
   phenomenon and the conclusions \cite{2,19} on rather a rapid growth of
   $\alpha$ in the previous epoch. The field $\sigma$ is also able to play
   the role of the so-called quintessence explaining the Universe
   acceleration observed in modern cosmology.

   Other predictions of Fujii's theory are the existence of a non-Newtonian
   short-range component of gravity and a weak equivalence principle
   violation. The orders of magnitude of both effects are compatible with
   modern experimental constraints.

\subsection{Other models}

   We have briefly described two sufficiently well-known and elaborated
   phenomenological theories. Let us also mention some alternative
   approaches.

   Anchordoqui and Goldberg \cite{35} discussed the so-called quintessence,
   i.e., a minimally coupled scalar field playing the role of an
   ``accelerator'' of the Universe expansion, as a possible generator of
   $\alpha$ variability. They considered scalar field potentials combining a
   number of exponential functions with different coefficients. Like Fujii's
   theory, this model makes it possible to reconcile the tough constraints
   obtained from the Oklo reactor with the results of Webb et al. obtained
   from quasar absorption spectra: they become compatible due to slowing
   down of the scalar field evolution in approaching the present epoch.

   Damour and Polyakov \cite{36} suggest a more general approach: among the
   models of the Universe containing ordinary matter and a real scalar
   field, one singles out a class of low-energy models possessing invariance
   under diffeomorphisms and gauge transformations. The resulting theories
   have a Born-Infeld type Lagrangian structure. In such theories, the
   masses and $g$-factors of fermions, as well as their electromagnetic
   coupling constants, are scalar field dependent, whereas their electric
   charges and gravitational coupling may remain constant. One then analyzes
   the effect of all variable factors on the observed atomic spectra. The
   latter, as follows from Ref.\,\cite{36}, bear information on the possible
   variability of $\alpha$ as well as the $g$-factors and particle mass
   ratios, and, moreover, all predictions concerning the atomic spectra turn
   out to be conformal gauge independent.

   Bertolami et al. \cite{bert1}, among others, stress the effective Lorentz
   and CPT symmetry violation that accompanies variable fundamental constants
   in any theory where these constants are scalar field dependent since
   variable scalar fields inevitably create preferred space-time directions.
   As en example of such a theory, they analyze N=4 supergravity
   in four dimensions, containing two scalar fields, and show that
   some of its solutions lead to $\alpha(t)$ dependence compatible with the
   astrophysical data, though at the expense of some fine tuning.
   In another paper \cite{bert2}, Bento, Bertolami and Santos
   considered a cosmological model with a two-field quintessence,
   i.e., a set of two scalar fields $\phi$ and $\psi$, minimally coupled to
   the Einstein gravity and possessing the potential
\bear
      V(\phi,\psi) = \e^{-\lambda\psi} P(\phi,\psi),
\ear
   where $P(\phi,\psi)$ is a third-order polynomial depending on both
   fields, while the electromagnetic field Lagrangian had the form
\beq
      L_{\rm e-m} = \fract{1}{4}B(\phi,\psi) F\MN F\mn,
\eeq
   $B(\phi,\psi)$ being a linear function of both scalars. It was found that
   some solutions for this system describe a cosmology with a transient late
   period of accelerated expansion, making it possible to fit the data
   arising from quasar absorption spectra and comply with the bounds on the
   variation of $\alpha$ from the Oklo reactor, meteorite analysis, atomic
   clock measurements, cosmic microwave background radiation, and big bang
   nucleosynthesis.

   Kirillov (see \cite{kirv05} and references therein) suggested a modified
   field theory which is aimed at explaining all effects ascribed to dark
   matter and dark energy without explicitly introducing them. A
   phenomenological manifestation of this theory may be expressed as a
   modification of the law of gravity, such that, for a point source
   of mass $M_0$, the gravitational potential reads
\beq
      \phi = - \frac{GM_0}{r}[1 + f(r)],
\eeq
   where the function $f(r)$ may be chosen in the form
\beq
      f(r) = \vars{ (r/r_0) \ln (r_{\max}\e/r), \cm & r \leq r_{\max},\\
             r_{\max}/r_0,                & r > r_{\max};
          }
\eeq
   the parameter $r_0$, having the order of 1--5 kpc, is different for
   different galaxies while $r_{\max}$ only depends on cosmological time
   and is now of the order of 100 Mpc. At $r > r_{\max}$, Newton's law is
   restored but with an effective mass $M_{\max} = M_0(1+ r_{\max}/r_0)$.
   It is shown that such a modification is able to account for all dark
   matter effects observed and, in particular, to explain the whole variety
   of ``dark matter halos'' in astrophysical systems. It is also shown that,
   in the modified theory, the galaxy formation process leads to an
   effective equaton of state of the fictituous ``dark matter'' with $w <
   -1/3$ ($w$ being the ratio of pressure to energy density) and therefore
   it can play the role of dark energy. The standard picture of the early
   Universe is preserved, but the theory predicts variation of all
   interaction constants (including $G$ and $\alpha$) with the same time
   dependence \cite{kirv04}.

\section {Concluding remarks.
    Physical and metrological significance of FPC variations}

   General theoretical considerations related to the necessary unification
   of all interactions and the requirement of unity of the physical science
   lead, probably inevitably, to the idea of a dynamical nature of all known
   FPCs or at least a greater part of them. Thus more surprising is that so
   much effort is needed to discover their variability than the belief that
   they are variable. Meanwhile, the experimental confirmations of these
   ideas are at present very poor and unreliable. This certainly concerns
   all FPCs, not only $\alpha$.

   An important and natural feature of all models is the interrelation of
   the predicted variations of different FPCs. Therefore, in both planning
   and interpretation of future experiments, such interrelations should be
   taken into account. Thus, in the context of string theories, it turns out
   \cite{36} that the quantum chromodynamics constant $\Lambda_{\rm QCD}$
   and the weak interaction constant $v$ are ``even more variable'' than
   $\alpha$ \cite{37}:
\beq
    \Delta\Lambda_{\rm QCD}/\Lambda_{\rm QCD}
        \sim 30\Delta\alpha/\alpha, \cm
                        \Delta v/v \sim 80 \Delta\alpha/\alpha.
\eeq

   According to \cite{38}, any theoretical model, predicting cosmological
   variations of $\alpha$ over $10^{-6}$ due to changes of some long-range
   scalar field, also predicts violations of the weak equivalence principle
   (WEP) over $10^{-13}$. To discover such variations, it is only necessary
   to gain one order of magnitude as compared to the modern experimental
   constraints. This endows special significance to the space experiments
   for testing the WEP planned for the nearest years: MICROSCOPE
   (planned sensitivity $10^{-15}$), STEP (up to $10^{-18}$ \cite{40}) and
   others.

   Thus new space experiments and observations in the area of gravitation,
   cosmology and astrophysics (improved satellite observations of the
   microwave background, analysis of supernovae and quasar radiation, space
   experiments for testing the WEP and constancy of the gravitational
   constant $G$, laboratary tests of FPC stability etc.) are expected to
   bring new results of major significance for the whole physical picture of
   the world. The discovery of a dynamical nature of a number of FPCs, as
   well as revision of the level of fundamentality for many of them, will
   inevitably lead to the corresponding revision in the foundations of
   theoretical metrology. Following the change in the general theoretical
   paradigm, the set of FPCs will also be revised: the parameters, whose
   dynamical nature will be proved, will lose their fundamental status,
   giving way to the basic constants of an underlying unified theory.

   If $\alpha$ does vary, a question of great interest for both physics and
   metrology is: which of the quantities in the definition (1) of $\alpha$,
   namely, $c$, $\hbar$, $e$ or $\eps_0$, are really varying.
   Different viewpoints may be found in the literature.

   The theories mentioned in Sec.\,3 modify the Maxwell equations and
   actually introduce a scalar field dependence of the effective electronic
   charge $e$. On the other hand, Hehl and Obukhov \cite{HOb04} note that,
   since $e$ and $h$, as the units of charge and action, are invariants (4D
   scalars), to reconcile a variable $\alpha$ with the
   invariable (premetric) form of the Maxwell equations one may choose the
   speed of light $c$ to vary. For variable $c$ theories see, e.g.,
   Ref.\,\cite{peres}, the brief review \cite{36a} and references therein.
   Though, as noted by Alfonso-Faus \cite{faus}, such first principles as
   local Lorentz invariance and local position invariance disfavour
   variations of $c$ as compared to changes of $e$ as a reason for varying
   $\alpha$.

   Hehl and Obukhov \cite{HOb04} mention one more possibility: they
   represent $\alpha$ as
\beq                                                       \label{H-O}
      \alpha = e^2 \Omega_0/(4\pi \hbar) = \Omega_0/(2 R_K),
\eeq
   where $\Omega_0$ is the impedance of free space and $R_K$ is the von
   Klitzing constant (quantum Hall resistance). In these expressions, the
   speed of light $c$ has disappeared. If, again, $e$ and $\hbar$ are
   regarded real constants, the putative $\alpha$ variability should be
   attributed to the vacuum impedance $\Omega_0$ that becomes a dynamic
   (dilaton) field.

   Tobar \cite{Tobar} represented $\alpha$ in terms of the ratio of the
   quanta of electric ($\Phi_e$) and magnetic ($\Phi_m$) fluxes of force of
   the electron:
\beq                                                 \label{Tob-rep}
	\alpha = \frac{1}{4\sqrt{2}}
		\biggl(\frac{\Phi_e}{\Phi_m}\biggr)^{1/2}.
\eeq
   The quantity $\alpha$, being dimensionless, is independent of the system
   of units used. Eq.\,(\ref{Tob-rep}) makes this independence manifest,
   which is an advantage as compared with the conventional expressions (1).
   The representation (\ref{Tob-rep}) made it possible \cite{Tobar} to
   interpret the range variation of $\alpha$ at the electroweak energy scale
   (see the end of Sec. 2.1) as appearing due to equal components of
   electric screening (polarization of vacuum) and magnetic anti-screening
   (magnetization of vacuum), which cause the perceived quanta of electric
   charge to increase at small distances, while the magnetic flux quanta
   decrease. Thus there emerge the concepts of ``bare magnetic flux quanta''
   and ``bare electric charge''. With regard to a putative drift at
   astrophysical or cosmological scales, Tobar \cite{Tobar} also interpreted
   it as a differential drift of the electric and magnetic fluxes, but
   concluded that it is impossible to determine which fundamental constant
   is varying if $\alpha$ varies.

   In our view, the constants $c$, related to the space-time structure, and
   $\hbar$, related to the quantum properties of all kinds of matter, are
   more universal than $e$ that characterizes the electromagnetic
   interaction only and are therefore more likely to be true constants. It
   is quite natural that unification theories modify the Maxwell equations
   by introducing slowly varying scalars which make the effective electronic
   charge vary. This view is also consistent with Hehl-Obukhov's and Tobar's
   interpretations. On the other hand, the constant $\eps_0$, which is
   present in the SI definition of the fine structure constant, only
   connects the electric SI units with the mechanical ones and does not
   have a fundamental meaning. (See also a recent discussion of SI and
   other systems of units in Refs.\,\cite{borde,kon05}.)

   Thus a variable $\alpha$ raises a number of questions in fundamental
   metrology, requiring further experimental and theoretical studies.

   As to the needs of practical metrology, the existing constraints on
   possible variations of $\alpha$ in the modern epoch, no more than a
   few units of the 17th significant digit per year [see \eq (9)], compared
   to the achieved measurement accuracy of $10^9$, give no ground to
   expect any appreciable change in this constant in the foreseeable future.
   Thus there is no evident reason for taking into account the conjectural
   dynamical nature of $\alpha$ in the values of references of any physical
   quantities, their construction and analysis. However, as follows from
   our discussion, there are at least two exceptions: (i) observational and
   theoretical cosmology and astrophysics of remote quasars and galaxies,
   and (ii) high energy physics related to unification of interactions.

   A problem of importance for both theoretical and practical metrology is
   the connection between local and cosmological variations of the FPCs,
   $\alpha$ in particular. Such a connection should be known for
   comparison between, e.g., terrestrial and quasar measurements of
   $\alpha$ and $\dot\alpha$ and their proper interpretation.
   On the other hand, since this connection appears to be model-dependent
   \cite{shaw05,mota04a,mota04b}, spatial variations of FPCs deserve special
   attention due to their potential use for model
   selection in physics and cosmology.

\subsection*{Acknowledgment}

The authors are grateful to L.K. Isaev, V.N. Melnikov and M.I. Kalinin for
valuable discussions. K.B. thanks David Mota for helpful correspondence.

\end{document}